\pdfoutput=1
\documentclass[12pt,english]{aastex}
\usepackage[latin9]{inputenc}
\usepackage{geometry}
\usepackage[active]{srcltx}
\usepackage{amsbsy}
\usepackage{amstext}
\usepackage{amssymb}
\usepackage{graphicx}
\usepackage{esint}

\makeatletter

\providecommand{\tabularnewline}{\\}

\usepackage{amsbsy}
\usepackage{babel}
\newcommand{\mathd}{\mathrm{d}}
\newcommand{\mathe}{\mathrm{e}}

\makeatother

\usepackage{babel}
\begin{document}

\title{Effects of anisotropy of turbulent convection in mean-field solar
dynamo models}

\author{V.V. Pipin$^{1-3}$ and A.G. Kosovichev$^{3}$}

\affil{$^{1}$Institute of Solar-Terrestrial Physics, Russian Academy of
Sciences, \\
 $^{2}$ Institute of Geophysics and Planetary Physics, UCLA, Los
Angeles, CA 90065, USA \\
 $^{3}$Hansen Experimental Physics Laboratory, Stanford University,
Stanford, CA 94305, USA }
\begin{abstract}
We study how anisotropy of turbulent convection affects diffusion
of large-scale magnetic fields and the dynamo process on the Sun.
The effect of anisotropy is calculated in a mean-field magneto-hydrodynamics
framework using the minimal $\tau$-approximation. We examine two
types of mean-field dynamo models: the well-known benchmark flux-transport
model, and a distributed-dynamo model with the subsurface rotational
shear layer. For both models we investigate effects of the double-cell
meridional circulation, recently suggested by helioseismology. We
introduce a parameter of anisotropy as a ratio of the radial and horizontal
intensity of turbulent mixing, to characterize the anisotropy effects.
It is found that the anisotropy of turbulent convection affects the
distribution of magnetic fields inside the convection zone. The concentration
of the magnetic flux near the bottom and top boundaries of the convection
zone is greater when the anisotropy is stronger. It is shown that
the critical dynamo number and the dynamo period approach to constant
values for the large anisotropy parameter. The anisotropy reduces
the overlap of the toroidal fields of subsequent cycles in the time-latitude
``butterfly'' diagram. If we assume that sunspots are formed in
the vicinity of the subsurface shear layer, then the distributed dynamo
model with anisotropic diffusivity satisfies the observational constraints
from heloseismology results and is consistent with the value of effective
turbulent diffusion, estimated from the dynamics of surface magnetic
fields. 
\end{abstract}

\section{Introduction}

It has been long assumed that turbulent magnetic diffusion (or eddy
magnetic diffusivity) is an important part of the hydromagnetic dynamo
process on the Sun \citep{P55}. It transfers the energy of large-scale
magnetic fields to small scales and determines the characteristic
scale of the exited dynamo modes ( see, \citealp{park1971ApJ}, and
reviews of \citealp{2005PhR...417....1B}; and \citealp{chrev05}).
The magnitude of turbulent diffusion impacts the period of the dynamo
cycle, and the anisotropy of turbulent diffusion affects drifts of
the large-scale magnetic field during the activity cycle (e.g., \citealt{yosh1975,park,2000PhRvE..61.5202R,k02}).

Numerical simulations showed that the turbulent convection on the
Sun is anisotropic \citep{miesch08}. This anisotropy results from
influence of the global rotation on convective motions. It was shown
that convective motions form ``banana''-like giant cells along meridians.
In this case, the turbulent diffusivity in the meridional direction
exceeds the diffusivity in the radial direction (along the gravity
vector). \citet{k02} showed that such anisotropy brings the modeled
propagation of solar dynamo waves in better agreement with observations.
This fact was extensively used in various solar dynamo models \citep{kit:00,k02,pk11,pi13r}.

However, properties of the anisotropic magnetic eddy diffusivity,
which depends on the impact of the solar rotation on convective motions,
remains uncertain. Results of theoretical calculations of magnetic
turbulent diffusivity coefficients strongly depend on assumed models
of background turbulent flows. Mean-field magneto-hydrodynamics calculations
show that anisotropy of the diffusivity coefficients is strong for
the regime of fast rotation, when the Coriolis number $\Omega^{\star}=2\Omega_{0}\tau_{c}\gg1$;
here $\Omega_{0}$ is the angular velocity, and $\tau_{c}$ is a typical
convective turnover time. This regime indeed can be found in the lower
part of the solar convection zone. In the upper part, $\Omega^{\star}\le1$,
and the anisotropy is small. {Furthermore, the numerical simulations,
which are based on the test field method, (see, e.g., \citealt{2009A&A...500..633K,bran2012AA}),
confirm the analytical calculations of the rotation-induces anisotropy
effects in the mean electromotive force, including the coefficients
of the magnetic turbulent diffusivity. The global numerical simulations
reveal a strong anisotropy of convection in the upper part of the
convection zone, where, $\Omega^{\star}\le1$, (see, e.g., \citealt{miesch08,racin2011},\citealt{2013arXiv1301.1330G}).
Similar results are suggested by the nonlocal stellar convection theory
\citep{chan06}.The origin of this effect is unclear currently. This
anisotropy may self-consitently appear with the subsurface shear layer,
which is generated in the models (see, e.g. \citealt{miesch08,2013arXiv1301.1330G}). }

In theoretical dynamo calculations this fact can be taken into account
if we introduce an additional parameter to model the anisotropy of
the background turbulent flows in terms of the relative difference
of RMS velocity fluctuations of radial and horizontal flow components
(Eq. \ref{eq:anis}). Such approach has already been used in a mean-field
model of solar differential rotation \citep{kit2004AR,kit-r11}. In
particular, the anisotropy allowed to explain the subsurface shear
of the solar angular velocity. In our paper, we extend this idea and
compute the magnetic diffusivity tensor for a range of the parameter
of anisotropy. The calculations are performed using the so-called
minimal $\tau$ approximation of the mean field magneto-hydrodynamics
\citep{bf:02,2003GApFD..97..249R,2005PhR...417....1B}. Having in
mind the previous results by \citet{park1971ApJ} and \citet{k02},
we expect that the anisotropy due to additional horizontal diffusion
of magnetic field changes the direction of the dynamo wave propagation
and increases the horizontal scale of the mean magnetic field. We
find that this effect decreases the overlap between the ``butterfly
wing'' of the time-latitude diagrams evolution of the large-scale
toroidal magnetic field, improving agreement of the dynamo model with
observations.

The paper is structured as follows. In the next section we shortly
outline the basic equations and assumptions. Next, we examine the
simplified bechmark model suggested by \citealt{2008A&A...483..949J}
and investigate the anisotropy effects in more detailed mean-field
models \citep{pip2013ApJ,PK13}, which include the
subsurface rotational shear layer and the double-cell meridional circulation,
which was suggested by recent helioseismology results. In section
3 we summarize the main results. Some mathematical details are given
in Appendix.

\section{Basic equations}

We decompose the flow $\mathbf{U}$ and magnetic field $\mathbf{B}$
into the sum of the mean and fluctuating parts: $\mathbf{U}=\overline{\mathbf{U}}+\mathbf{u}$,
$\mathbf{B}=\overline{\mathbf{B}}+\mathbf{b}$; $\overline{\mathbf{U}}$,
$\overline{\mathbf{B}}$ represent the mean large-scale fields. Hereafter,
we use the small letters for the fluctuating parts of the fields and
capital letters with a over-bar for the mean fields. The mean effect
of the fluctuating turbulent flows and magnetic fields on the large-scale
magnetic field is described by the mean electromotive force, $\mathbf{\mathcal{E}}=\mathbf{\overline{u\times b}}$,
where the averaging is performed over an ensemble of the fluctuating
fields. Following the two-scale approximation \citep{rob-saw,KR80}
we assume that the mean fields vary over much larger scales (both
in time and space) than the fluctuating fields. The governing equations
for fluctuating magnetic field and velocity are written in a rotating
coordinate system as follows 
\begin{eqnarray}
\frac{\partial\mathbf{b}}{\partial t} & = & \boldsymbol{\nabla}\times\left(\mathbf{u}\times\mathbf{\overline{B}}\right)+\eta\nabla^{2}\mathbf{b}+\mathbf{\mathfrak{G}},\label{induc}\\
\frac{\partial u_{i}}{\partial t}+2(\Omega\times u)_{i} & = & -\nabla_{i}\left(p+\frac{(\mathbf{B\cdot b})}{\mu_{0}}\right)+\nu\Delta u_{i}+\frac{1}{\mu_{0}}\nabla_{f}(\bar{B}_{f}b_{i}+\bar{B}_{i}b_{f})+f_{i}+\mathfrak{F}_{i},\label{fluc}
\end{eqnarray}
where $\mathfrak{G}$ and $\mathfrak{F}$ denote the nonlinear contributions
of fluctuating fields, $p$ is the fluctuating pressure, $\boldsymbol{\mathbf{\Omega}}$
is the angular velocity, $\mathbf{f}$ is a random force driving the
turbulence. Equations (\ref{induc}) and (\ref{fluc}) are used to
compute the mean electromotive force, $\mathbf{\mathcal{E}}$. Details
of the calculations are given in Appendix.

It is known that rotation quenches the magnitude of the turbulent
diffusivity and induces the anisotropy of diffusivity along the rotation
axis \citep{kit-pip-rud,2008ApJ...676..740B}. Similar quenching effect
exists for the anisotropic background turbulent flows (see, Eqs\ref{eqs:emfA}-\ref{eq:phi3}).
We found that both anisotropic and isotropic parts of the turbulent
diffusivity are almost equally affected (quenched) by rotation. Therefore,
their ratio does not depend on the Coriolis number. In a simple case,
when we disregard the effect of the Coriolis force the magnetic diffusion
can be written as follows: 
\begin{equation}
\boldsymbol{\mathcal{E}}^{(dif)}=-\eta_{T}\boldsymbol{\nabla}\times\mathbf{\overline{B}}-\frac{a}{2}\eta_{T}\left(\boldsymbol{\nabla}-\mathbf{g}\left(\mathbf{g}\cdot\nabla\right)\right)\times\overline{\mathbf{B}},\label{eq:emf0}
\end{equation}
where the first term describes the isotropic diffusion, and the second
term describes the anisotropy. The parameter, $a$, quantifies the
level of the anisotropy (Eq.\ref{eq:anis}). In the general case the
expression for the anisotropic part of diffusion is given by Eq.(\ref{eq:result}).
The mixing-length theory of stellar convection requires that $0\le a\le4$
(see, \citealt{kit2004AR,2005A&A...431..345R}). Numerical simulations
indicate higher values up to $a\sim10$ (see, e.g., Fig. 13 in \citealt{miesch08}).

In the paper we study the standard mean-field induction equation in
perfectly conductive media: 
\begin{equation}
\frac{\partial\overline{\mathbf{B}}}{\partial t}=\boldsymbol{\nabla}\times\left(\mathbf{\boldsymbol{\mathcal{E}}+}\overline{\mathbf{U}}\times\overline{\mathbf{B}}\right)\label{eq:dyn}
\end{equation}
where $\boldsymbol{\mathcal{E}}$ is the mean electromotive force;
$\overline{\mathbf{U}}=\mathbf{e}_{\phi}r\sin\theta\Omega\left(r,\theta\right)+\overline{\mathbf{U}}^{p}\left(r,\theta\right)$
is the mean flow which includes the differential rotation, $\Omega\left(r,\theta\right)$,
and meridional circulation, $\overline{\mathbf{U}}^{p}\left(r,\theta\right)$;
the axisymmetric magnetic field is given: 
\[
\overline{\mathbf{B}}=\mathbf{e}_{\phi}B+\nabla\times\frac{A\mathbf{e}_{\phi}}{r\sin\theta}
\]
where $r$ is radius and $\theta$ - polar angle, $B\left(r,\theta\right)$
is the strength of the toroidal component of magnetic field, $A\left(r,\theta\right)$
represents vector potential of the poloidal component.

\begin{table}

\caption{Benchmark models design and parity preference\label{tabl1}}

\begin{centering}
\begin{tabular}{|c|c|c|c|c|c|c|c|}
\hline 
 & $\alpha$-effect & B-L term & \multicolumn{3}{c|}{Circulation } & \multicolumn{2}{c|}{Parity}\tabularnewline
\cline{4-8} 
Model &  &  & $c_{1}$ & $c_{2}$ & $c_{3}$ & a=0 & a=4\tabularnewline
\hline 
B & + & - & - & - & - & A & S\tabularnewline
\hline 
C1 & - & + & 1 & - & - & A & A\tabularnewline
\hline 
C2 & - & + & 0.5 & 1.5 & - & S & S\tabularnewline
\hline 
C3 & - & + & 0 & 1 & 2.5 & A & A\tabularnewline
\hline 
\end{tabular}
\par\end{centering}

\end{table}

\subsection{Benchmark models design}

\begin{figure}
\begin{centering}
\includegraphics[width=0.95\columnwidth]{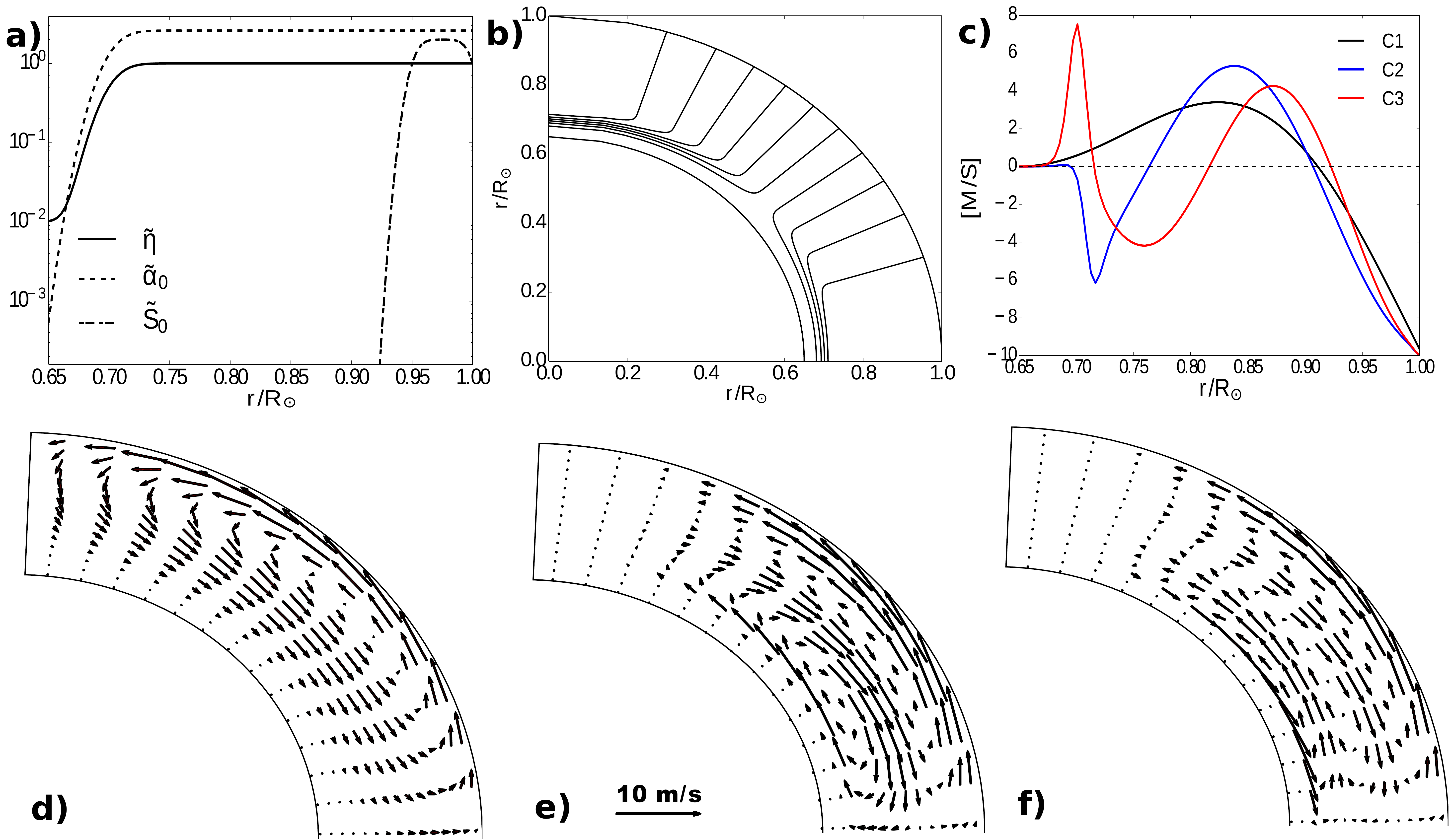}
\par\end{centering}

\label{fig:mc}Parameters of the benchmark model: a) the radial profiles
of the turbulent diffusivity, the $\alpha$-effect and the Babcock-Leighton
generation term; b) the angular velocity distribution; c) the radial
profiles of the latitudinal velocity field at $30^{\circ}$ latitude;
d) the velocity field for the model C1; e) and f) the same for the
models C2 and C3
\end{figure}

\begin{figure}
\begin{centering}
\includegraphics[width=0.45\textwidth]{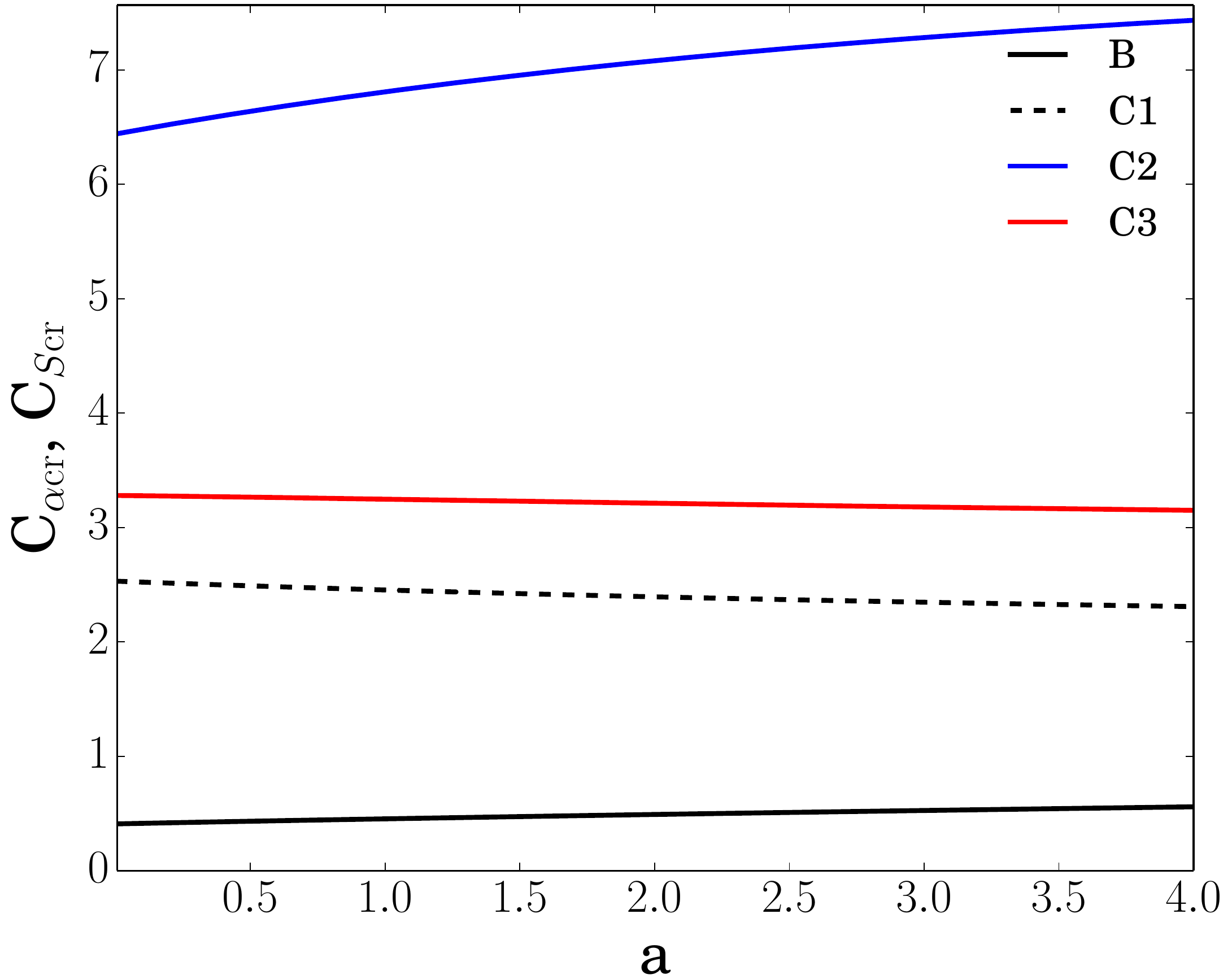}
\includegraphics[width=0.45\textwidth]{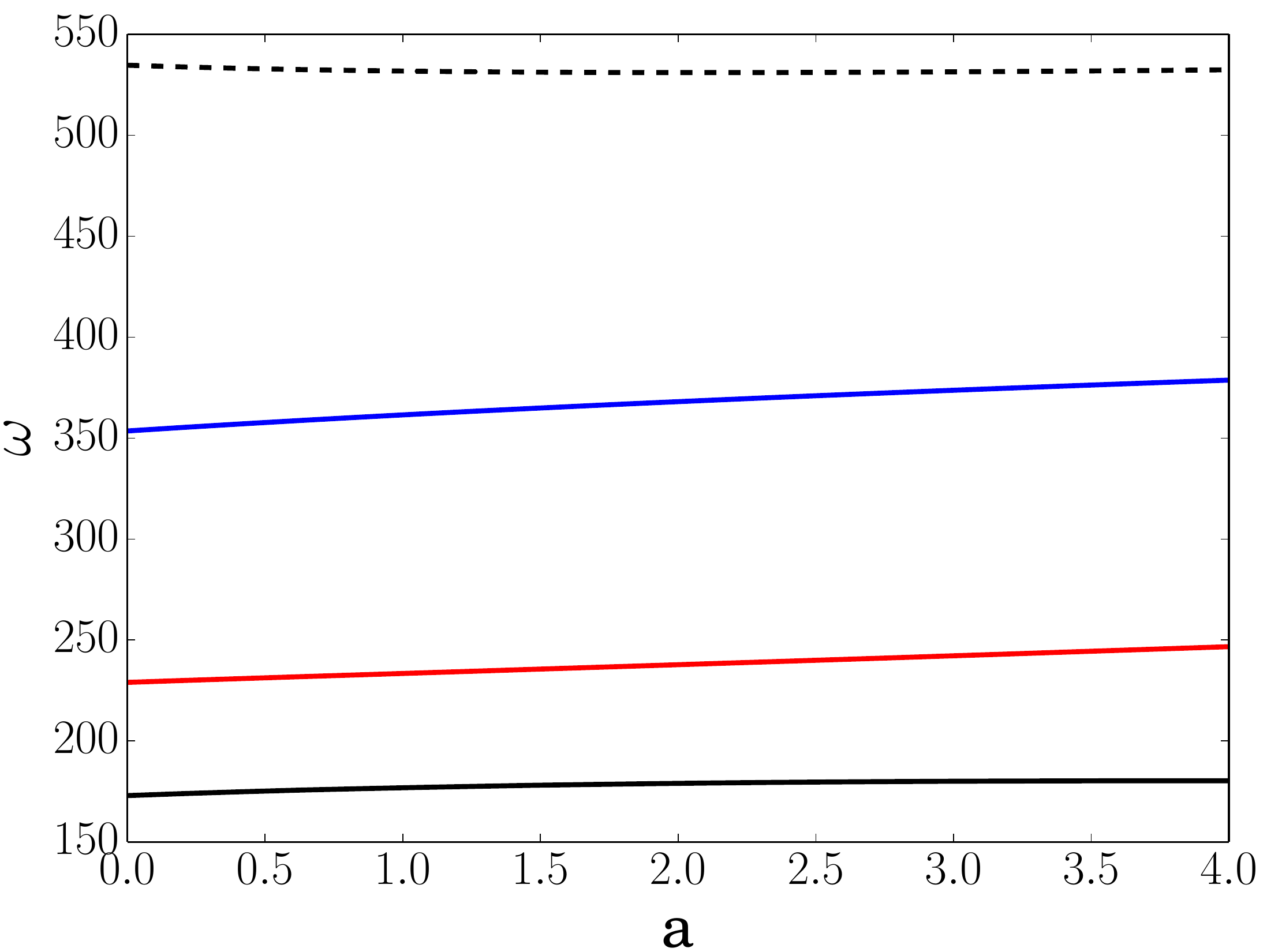} 
\par\end{centering}

\caption{\label{fig:param}Panel a) shows the dynamo instability threeshold
for different models. For the model B it shows the critical parameter
$C_{\alpha}$ and for the model C1 and C2 it shows critical parameter
$C_{S}$ ; b) shows the frequency of the first unstable dipole- and
quadrupole-like modes;}
\end{figure}

{In this section we examine the effect of the anisotropic mixing
using the benchmark model presented by \citet{2008A&A...483..949J}(hereafter
J08). In this case we use the simplest representation for the mean
electromotive force with the isotropic $\alpha$-effect, $\boldsymbol{\boldsymbol{\mathcal{E}}}=\alpha\overline{\mathbf{B}}+\boldsymbol{\boldsymbol{\mathcal{E}}}^{(dif)}$,
where $\boldsymbol{\boldsymbol{\mathcal{E}}}^{(dif)}$ is given by
Eq.(\ref{eq:emf0}). Thus, we have the following dynamo equations:}

{
\begin{eqnarray*}
\frac{\partial A}{\partial T} & = & \tilde{\eta}\frac{\partial^{2}A}{\partial x^{2}}+\tilde{\eta}\left(1+\frac{a}{2}\right)\frac{\sin\theta}{x^{2}}\frac{\partial}{\partial\theta}\frac{1}{\sin\theta}\frac{\partial A}{\partial\theta}\\
 & + & \frac{\tilde{\eta}a}{2x}\frac{\partial A}{\partial x}-R_{e}\left(\frac{\tilde{U}_{\theta}}{x}\frac{\partial A}{\partial\theta}+\tilde{U}_{r}\frac{\partial A}{\partial x}\right)\\
 & + & C_{\alpha}\tilde{\alpha}_{0}x\cos\theta\sin^{3}\theta B+C_{S}\tilde{S}_{0}x\cos\theta\sin\theta B\left(0.7R_{\odot},\theta,T\right)\\
\frac{\partial B}{\partial T} & = & \frac{C_{\Omega}}{x}\left(\frac{\partial\tilde{\Omega}}{\partial x}\frac{\partial A}{\partial\theta}-\frac{\partial\tilde{\Omega}}{\partial\theta}\frac{\partial A}{\partial x}\right)+\frac{a}{2x}\frac{\partial}{\partial x}\left(\tilde{\eta}B\right)\\
 & + & \left(1+\frac{a}{2}\right)\frac{\tilde{\eta}}{x^{2}}\frac{\partial}{\partial\theta}\frac{1}{\sin\theta}\frac{\partial\sin\theta B}{\partial\theta}+\frac{1}{x}\frac{\partial}{\partial x}\tilde{\eta}\frac{\partial xB}{\partial x}\\
 & - & \frac{R_{e}}{x}\left(\frac{\partial\left(x\tilde{U}_{r}B\right)}{\partial x}+\frac{\partial\tilde{U}_{\theta}B}{\partial\theta}\right),
\end{eqnarray*}
where $C_{\alpha}={\displaystyle \frac{\alpha_{0}R^{3}}{\eta_{0}}}$,
$C_{\Omega}={\displaystyle \frac{\Omega_{0}R^{2}}{\eta_{0}}}$, $R_{e}={\displaystyle \frac{U_{0}R}{\eta_{0}}}$,
and $C_{S}$ is to control the amplitude of the Babcock-Leighton effect,
$x=r/R_{\odot}$, and $T=t{\displaystyle \frac{\eta_{0}}{R_{\odot}^{2}}}$,
where $\eta_{0}=10^{11}\textnormal{cm}^{2}\textnormal{s}^{-1}$ is
the background level of the magnetic turbulent diffusivity. The radial
profiles of the angular velocity, $\tilde{\Omega}$, the turbulent
diffusivity, $\tilde{\eta}$, the $\alpha$-effect, $\tilde{\alpha}_{0}$,
and the Babcock-Leighton effect, $\tilde{S}_{0}$ , are shown in Figure\ref{fig:mc}(a,b).
They are the same as in \citep{2008A&A...483..949J}( J08). The dynamo
domain of benchmark model is located between $r_{b}=0.65R_{\odot}$
and $r_{e}=R_{\odot}$. }

{The meridional flow is modeled in the form of stationary circulation
cells stacking cells along the radius. The pattern is modeled by the
stream functions $\Psi$:}

{
\begin{eqnarray}
\Psi & = & -\frac{2}{\pi}\frac{\left(x-x_{b}\right)^{2}}{1-x_{b}}\sin\left(\frac{\pi\left(x-x_{b}\right)}{\left(1-x_{b}\right)}\right)\sin\theta\cos\theta\label{eq:psi}\\
\Psi & = & \frac{2c_{0}}{\pi\left((1+\exp(-200(x-x_{0})\right)}\left(\frac{1-x}{x}\right)^{1.5}\left(\frac{\partial P_{2}}{\partial\theta}+m\frac{\partial P_{4}}{\partial\theta}\right)\times\sum_{n=1}^{3}c_{n}\sin\left(\frac{n\pi\left(x-x_{b}\right)}{\left(1-x_{b}\right)}\right)\label{eq:psim}
\end{eqnarray}
where, $P_{2,4}$ are the Legendre polynomials, $x_{b}=r_{b}/R$ is
the inner boundary of the integration domain; parameter $m$ controls
the number of cells in latitude; $c_{0}$ is the constant to normalize
the maximum of the flow amplitude to 1; $c_{1}$, $c_{2}$ and $c_{3}$control
the amplitudes of flows in the stacking cells. The velocity field
of the flow is given by $\overline{\mathbf{U}}^{p}={\displaystyle U_{0}\boldsymbol{\nabla}\times}\left(\mathbf{e_{\phi}\Psi}\right)$,
in the case of one cell circulation (model C1, Eq.\ref{eq:psi}) and
$\overline{\mathbf{U}}^{p}={\displaystyle \frac{U_{0}}{\overline{\rho}}\boldsymbol{\nabla}\times}\left(\mathbf{e_{\phi}\Psi}\right)$
for the models C2 and C3, (the case Eq.\ref{eq:psim}). The $U_{0}$
is a characteristic flow speed. In the cases C2 and C3 we choose $x_{0}=0.71$
to cut off penetration of circulation below $r=0.7R_{\odot}$. The
stream function is similar to the one of \citep{PK13} with a modification
to control the penetration of the meridional circulation below the
convection zone. The circulation pattern is illustrated in Figure
\ref{fig:mc}. We will use the same value for the $\Omega$-effect,
as in the J08, $C_{\Omega}=1.4\times10^{5}$. The parameters of the
benchmark models are listed in the Table\ref{tabl1}. }

Summarizing, we conclude that models B and C1 correspond to those
studied by J08. Models C2 and C3 are given for comparison. The case
C2 is motivated by recent results of helioseismology results \citep{Zhao13m}.
The radial profile of the latitudinal component of meridional circulation
(see, Fig.\ref{fig:mc}(c, blue curve)) and geometry of the flow Fig.\ref{fig:mc}(e))
are close to results detected by helioseismology inversion. {Numerical
simultaions often produce the multi-cellular meridional circulation
which can have three cells stacked along the radial direction \citep{kap2012,2013arXiv1301.1330G}.
This question is addressed for the models with triple-cell circulation
pattern (Fig.\ref{fig:mc}(c, red curve) and Fig.\ref{fig:mc}(f)).
In this case our circulation pattern is only qualitatively reproduce
the numerical simulations. The study of this case help us to highlight
the important difference between the models for the case of the odd
and even number of circulation cells stacked along the radius. }

\begin{figure}
\includegraphics[width=0.97\textwidth]{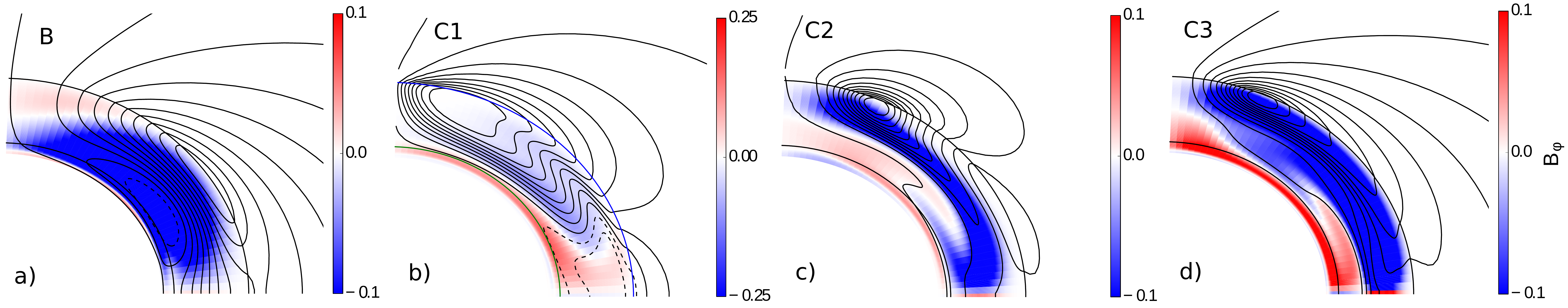}

\caption{\label{fig:snaps}{Panel a) shows a snapshot of the large-scale
magnetic field inside the convection zone for the benchmark dynamo
model B with $a=4$. The field lines show of the poloidal component
of the mean magnetic field, and the toroidal magnetic field is shown
by the background images. Panel b) shows the same for the model C1;
c) shows the same as b) for the model with the double-cell meridional
circulation, the model C2; d) shows the same for the model C3 and
$a=4$.}}
\end{figure}

\subsection{Results for benchmark models}

Here, we show results of the eigen-value problem solution for the
benchmark models listed in the Table 1. The dynamo instability develops,
when a non-dimensional parameters, $C_{\alpha}$, which controls the
magnitude of the $\alpha$-effect, or, $C_{S}$, which controls the
strength of the Coriolis force acting on the flux-tube rising through
the solar convection zonet, exceeds the critical value. Figure \ref{fig:param}(a)
shows the critical threshold parameters $C_{\alpha cr}$ and $C_{Scr}$
for the dynamo instability of the dipole like modes as a function
of the anisotropy parameter, $a$. Figure \ref{fig:param}(b) shows
the frequency of the first unstable mode. The cases B and C1 correspond
to those studied by \citet{2008A&A...483..949J}(J08). In these cases
we have $C_{\alpha cr}=0.408$, $\omega=173$ and $C_{Scr}=2.53$
$\omega=534$ for $a=0$. This is in perfect agreement with J08. The
main result is that the critical threshold dynamo parameters, as well
as, the frequency (and period) of the dynamo oscillations vary rather
little with variation of $a$. Moreover, in the cases B and C2 (even
number of circulation cells), the dynamo threshold is slowly growing
with increasing of $a$. On the other hand, the cases C1 and C3 show
the slowly decreasing dynamo threshold with increasing of $a$. As
we have guessed previously, \citep{PK13spd}, there is a similarity
for the dynamo regimes operating with even or odd number of the circulation
cells stacking along the radius. Another interesting finding is that
the dynamo period is growing with the increasing number of circulation
cells (see also \citealt{CH13m}). This is not directly related to
the effect of the anisotropy of turbulent diffusion.

Figures \ref{fig:param}(a,b) show the threshold parameters for the
first unstable dipole-type eigen modes. The threshold parameters of
the quadrupole type modes vary in similar way. For the model B, the
first unstable dipole type mode has smaller $C_{\alpha cr}$ than
the first unstable quadrupole type mode for $a<2$. Both modes have
close frequencies. For the model C1 the first unstable dipole type
mode is preferable for the all range of $a$. Meanwhile the frequency
of the first unstable quadrupole type mode is as twice smaller than
the frequency of the first unstable dipole type mode.  In the model
C2 the first unstable mode has the quadrupole type symmetry for the
all range of $a$. The opposite is true for the model C3. The results
for the parity preference are listed in the Table 1.

\begin{figure}
\includegraphics[width=0.97\textwidth]{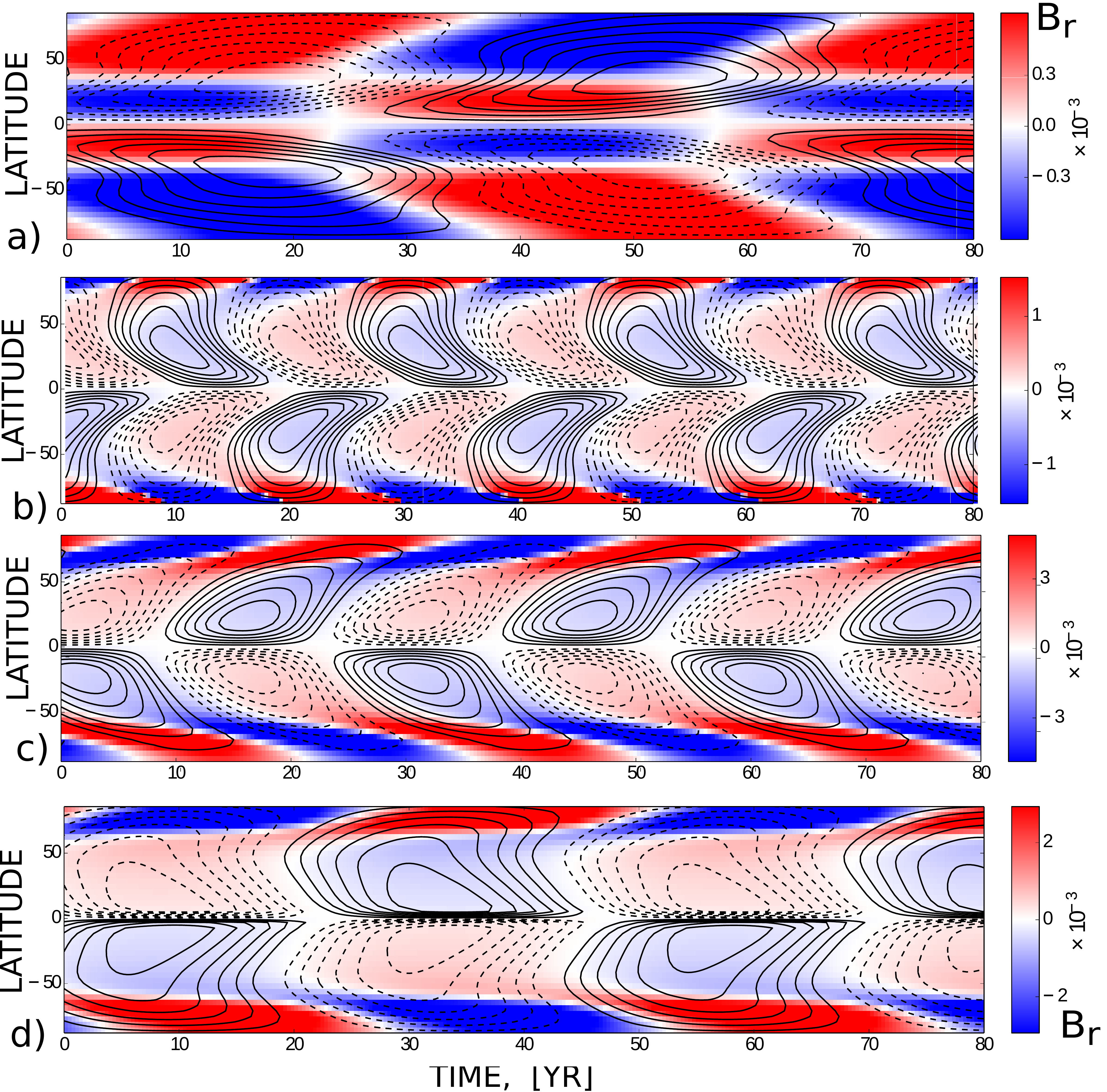}

\caption{\label{fig:bflyb}{Panel a) shows the time-latitude evolution
of the toroidal magnetic field at the bottom of the convection zone
(contours) and the radial magnetic field (background image) for the
benchmark dynamo model B for the anisotropy parameter $a=4$; b) shows
the same as (a) for the model C1; c) shows the same for the model
C2; d) shows the same for the model C3.}}
\end{figure}

{The typical snapshots of the magnetic field distributions
for the models B,C1,C2 and C3 for the case of $a=4$ are illustrated
in Figure \ref{fig:snaps}. The time-latitude diagrams of the toroidal
magnetic field at $r=0.7R_{\odot}$and the radial magetic field at
the surface are shown in Figure \ref{fig:bflyb}. The model C1 and
C3 have some qualitative agreement with observations. However, the
butterfly wings of the toroidal magnetic field are to wide in compare
with observations. Also, for the large anisotropy paprameter $a=4$
the model C1 has the wrong phase relation between the maxims of toroidal
magnetic field in equatorial region and inversion of the radial magnetic
field at the pole. The model C3 reproduces the phase relation in a
better way, though the inversion of the polar field occurs about 5
years in advance to the maximum of the toroidal field in equatorial
region. We have to note that for the case of $a=0$ the model C3 shows
much longer polar branch of the radial magnetic field. Thus, the including
the anisotropy of the turbulent diffusion in the model brings this
model in the better agreement with observations. In the models B and
C2 the toroidal field drifts to the pole at the bottom of the convection
zone. In the model C2, the meridional circulation moves the toroidal
field toward equator in the middle of the convection zone. Similar
to the models C1 and C3, in the models B and C2 the anisotropy of
turbulent diffusivity decrease the polar branch of the radial magnetic
field evolution. Additionally, in the model B the poleward drifting
dynamo wave of the toroidal magnetic field converge to the steady
wave near the surface.}

{Summarizing consideration of the benchmark models we conclude,
that the radial anisotropy of the turbulent diffusivity does not significantly
change the conditions for the dynamo instability. It does not impact
very much the dynamo period as well. However, we find that it can
result to the shorter polar branch of the radial magnetic field at
the surface. For the models B, C2 and C3, in the upper part the convection
zone the poleward migration of the toroidal magnetic field dominates.
This migration is reversed in the dynamo model with the subsurface
shear of the angular velocity, which we study in the next section.}

\begin{figure}
\includegraphics[width=0.96\textwidth]{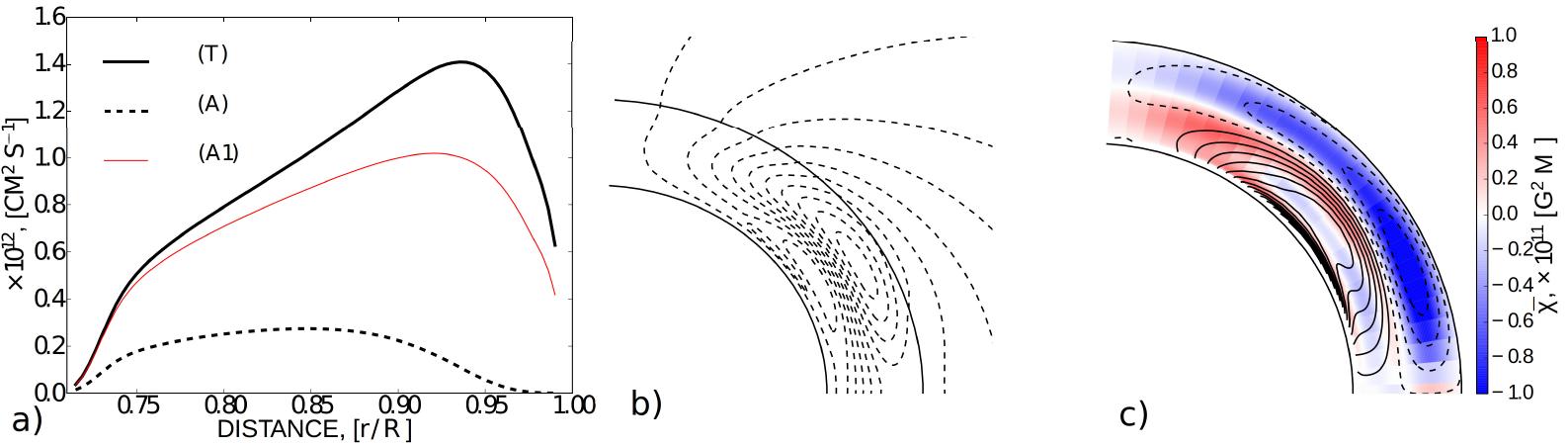}

\caption{\label{fig:prof}Panel a) Turbulent diffusion coefficients as a function
of radius: $\eta^{(A)}$ is the anisotropic part due to rotation,
$\eta^{(A1)}$is anisotropic diffusivity for $a=3$, $\eta^{(T)}$is
the total diffusivity. Panel b) snapshot of the poloidal magnetic
field lines; c) shows contours for the toroidal magnetic field (contours
$\pm1$kG) and the small-scale magnetic helicity is in the background. }
\end{figure}

\subsection{Solar dynamo model with subsurface shear}

{In this section we consider the anisotropy effects in the
model, which we developed in our recent papers \citep{pk11,PK13,pipea2012AA}.
The mean electromotive force is given as follows  (Pipin 2008, hereafter,
P08). 
\begin{equation}
\mathcal{E}_{i}=\left(\alpha_{ij}+\gamma_{ij}^{(\Lambda)}\right)\overline{B}_{j}-\left(\eta_{ijk}+\eta_{ijk}^{(\delta)}\right)\nabla_{j}\overline{B}_{k}+\mathcal{E}_{i}^{(A)}.\label{eq:EMF-1}
\end{equation}
where $\mathcal{E}_{i}^{(A)}$ is the anisotropic part of magnetic
diffusivity for the prescribed anisotropy of the backgroud turbulence
model. It is given by Eqs (\ref{eq:result}) and (\ref{eqs:emfA}).
The tensor $\alpha_{ij}$ describes the $\alpha$-effect. It includes
hydrodynamic ($\alpha_{ij}^{(H)}$) and magnetic ($\alpha_{ij}^{(M)}$)
helicity contributions: 
\begin{eqnarray}
\alpha_{ij} & = & C_{\alpha}\psi_{\alpha}\sin^{2}\theta\alpha_{ij}^{(H)}+\alpha_{ij}^{(M)}\label{alp2d}
\end{eqnarray}
The $\alpha$-quenching function $\psi_{\alpha}=-3/4\phi_{6}^{(a)}\left(\beta\right)$
depends on $\beta={\displaystyle \left|\overline{B}\right|/\sqrt{\mu_{0}\overline{\rho}\overline{u^{2}}}}$,
and $\phi_{6}^{(a)}$ is given in P08. The magnetic helicity contribution
to the $\alpha$-effect is defined as follows (P08): 
\begin{equation}
\alpha_{ij}^{(M)}=2\left(f_{2}^{(a)}\delta_{ij}-f_{1}^{(a)}\frac{\Omega_{i}\Omega_{j}}{\Omega^{2}}\right)\frac{\overline{\chi}\tau_{c}}{\mu_{0}\overline{\rho}\ell^{2}}\label{alpM}
\end{equation}
The functions $f_{1,2}^{(a)}$ describe the effect of rotation and
can be found in P08. The evolution of magnetic helicity $\overline{\chi}=\overline{\mathbf{a}\cdot\mathbf{b}}$,
where $\mathbf{a}$ is the fluctuating vector-potential, $\mathbf{b}$-
the fluctuating magnetic field is determined from the conservation
law (see, \citealp{pi13r,pip2013ApJ}):}

{
\begin{equation}
\frac{\partial\overline{\chi}^{(tot)}}{\partial t}=-\frac{\overline{\chi}}{R_{m}\tau_{c}}-\eta\overline{\mathbf{B}}\cdot\mathbf{\overline{J}}-\mathbf{\left(\overline{U}\cdot\boldsymbol{\nabla}\right)}\overline{\chi}^{(tot)}\label{eq:helcon}
\end{equation}
where $\overline{\chi}^{(tot)}=\mathbf{\overline{A}}\cdot\overline{\mathbf{B}}+\overline{\chi}$
is the total magnetic helicity. In the model we assume $R_{m}=10^{6}$. }

{The turbulent pumping coefficient in Eq(\ref{eq:EMF-1}),
$\gamma_{ij}^{(\Lambda)}$, depends on the mean density and turbulent
diffusivity stratification, and also on the Coriolis number $\Omega^{*}=2\tau_{c}\Omega_{0}$,
where $\tau_{c}$ is a typical convective turnover time, and $\Omega_{0}$
is the angular velocity. For detailed expressions of $\gamma_{ij}^{(\Lambda)}$
see the above cited papers. The turbulent diffusivity is anisotropic
due to the Coriolis force, and is given by: 
\begin{equation}
\eta_{ijk}=3\eta_{T}\left\{ \left(2f_{1}^{(a)}-f_{2}^{(d)}\right)\varepsilon_{ijk}-2f_{1}^{(a)}\frac{\Omega_{i}\Omega_{n}}{\Omega^{2}}\varepsilon_{njk}\right\} .\label{eq:diff}
\end{equation}
We also include the nonlinear effects of magnetic field generation
induced by the large-scale current and global rotation, which are
usually called the $\Omega\times J$-effect or the $\delta$ dynamo
effect \citep{rad69}. Their importance is supported by the numerical
simulations \citep{2008A&A...491..353K,2011A&A...533A.108S}. We use
the equation for $\eta_{ijk}^{(\delta)}$ which was suggested in P08
(also, see, \citealp{2004astro.ph..7375R}): 
\begin{equation}
\eta_{ijk}^{(\delta)}=3\eta_{T}C_{\delta}f_{4}^{(d)}\frac{\Omega_{j}}{\Omega}\left\{ \tilde{\varphi}_{7}^{(w)}\delta_{ik}+\tilde{\varphi}_{2}^{(w)}\frac{\overline{B}_{i}\overline{B}_{k}}{\overline{B}^{2}}\right\} ,\label{eq:delta}
\end{equation}
where, $C_{\delta}$ measures the strength of the $\Omega\times J$
effect, $\tilde{\varphi}_{2,7}^{(w)}\left(\beta\right)$ are normalized
versions of the magnetic quenching functions $\varphi_{2,7}^{(w)}$
given in P08. They are defined as follows, $\tilde{\varphi}_{2,7}^{(w)}\left(\beta\right)=\frac{5}{3}\varphi_{2,7}^{(w)}\left(\beta\right)$.
The functions $f_{\{1,2\}}^{(a,d)}$in Eqs (\ref{alp2d},\ref{eq:diff},
\ref{eq:delta}) depend on the Coriolis number. They can be found
in P08, as well.}

Following \citet{pk11apjl} we use a combination of the ``open''
and ``closed'' boundary conditions at the top, controlled by a parameter
$\delta=0.99$: 
\begin{equation}
\delta\frac{\eta_{T}}{r_{e}}B+\left(1-\delta\right)\mathcal{E}_{\theta}=0.\label{eq:tor-vac}
\end{equation}
This is similar to the boundary condition discussed by \citet{kit:00}.
This condition results to penetration of the toroidal field to the
surface, which increase the efficiency of the subsurface shear layer
\citep{pk11apjl}. For the poloidal field we apply a condition of
smooth transition from the internal poloidal field to the external
potential (vacuum) field.

Summing up, the model includes magnetic field generation generation
effects due to the differential rotation ($\Omega$ -effect), turbulent
kinetic helicity (the anisotropic $\alpha$-effect) and interaction
of large-scale currents with the global rotation, usually called $\Omega\times J$-effect
or $\delta$-effect \citep{rad69,2008A&A...491..353K,2011A&A...533A.108S}.
For the differential rotation, we use an analytical fit to the recent
helioseismology results of \citet{Howe2011JPh} (see, Fig.1(c) in
\citealp{PK13}). The subsurface rotational shear layer provides additional
energy for the toroidal magnetic field generation, and also induces
the equator-ward drift of the toroidal magnetic field \citep{pk11apjl}.
We also take into account the turbulent transport due to the mean
density and turbulent intensity gradients (so-called ``gradient pumping'').
The model includes also the magnetic helicity balance, as described
by \citet{pip2013ApJ} . The contribution of the anisotropic diffusion
to the mean electromotive force is given by Eqs(\ref{eqs:emfA}).

\begin{figure}
\includegraphics[width=0.8\columnwidth]{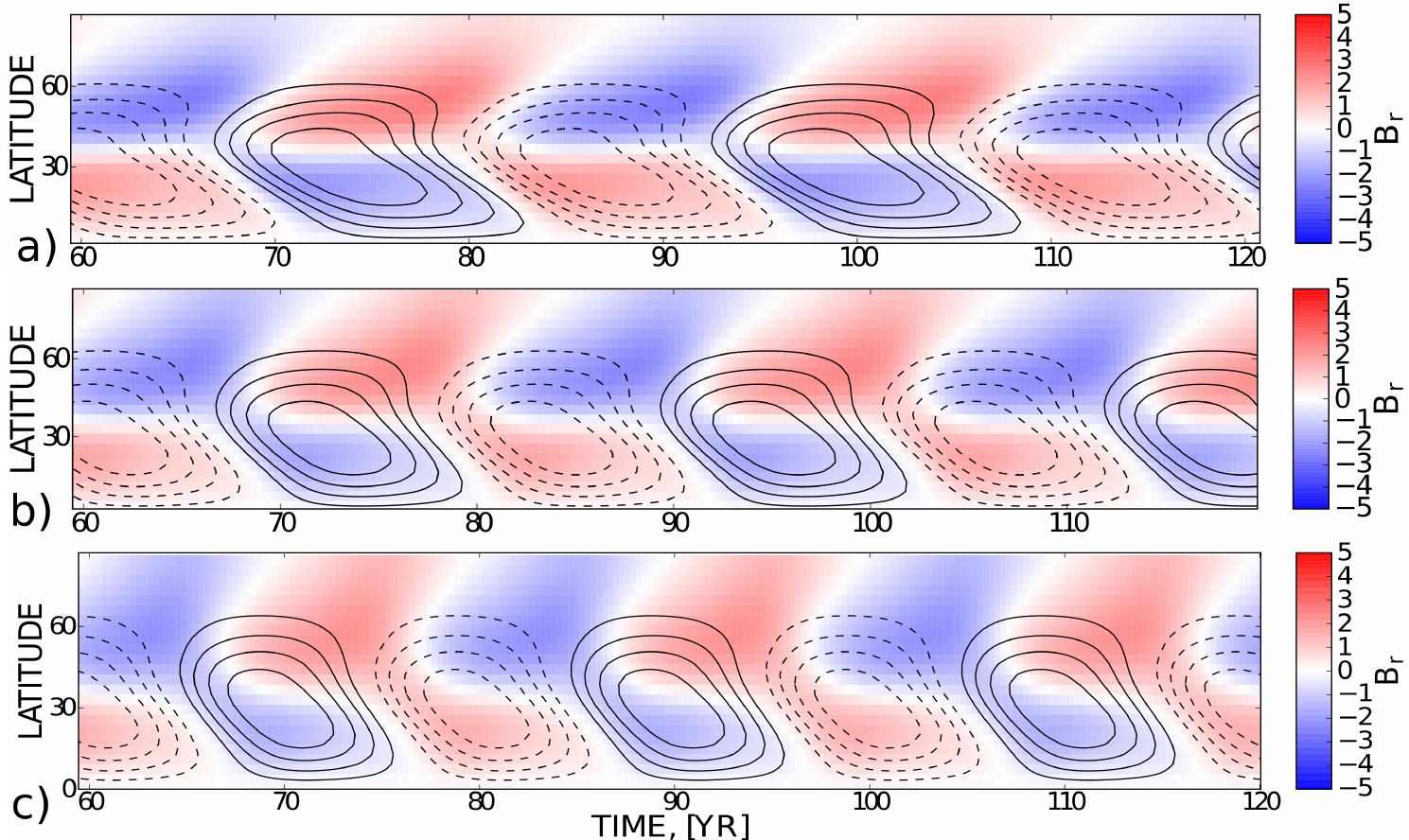}\caption{\label{fig:diag}(a) the time - latitude variations (``butterfly
diagrams'') for the model with $a=0$; (b) the same for the $a=2$
(c) the same for the $a=4$ . The toroidal field near the surface,
at $r=0.92R_{\odot}$, is shown by contours (plotted for $\pm100$G
range), and the surface radial magnetic field is shown by background
red-blue images.}
\end{figure}

The turbulent diffusivity profiles for $a=3$ are shown in Figure
\ref{fig:prof}(a). We see that with the anisotropy effect the total
effective diffusivity is greater than $10^{12}$ cm$^{2}$/s in the
upper part of the convection zone. This is close to the turbulent
magnetic diffusivity estimated from the sunspot decay rate \citep{1993A&A...274..521M}
and also from the cross-helicity observations \citep{2011SoPh..269....3R,2011ApJ...743..160P}.

\begin{figure}
\includegraphics[width=0.8\columnwidth]{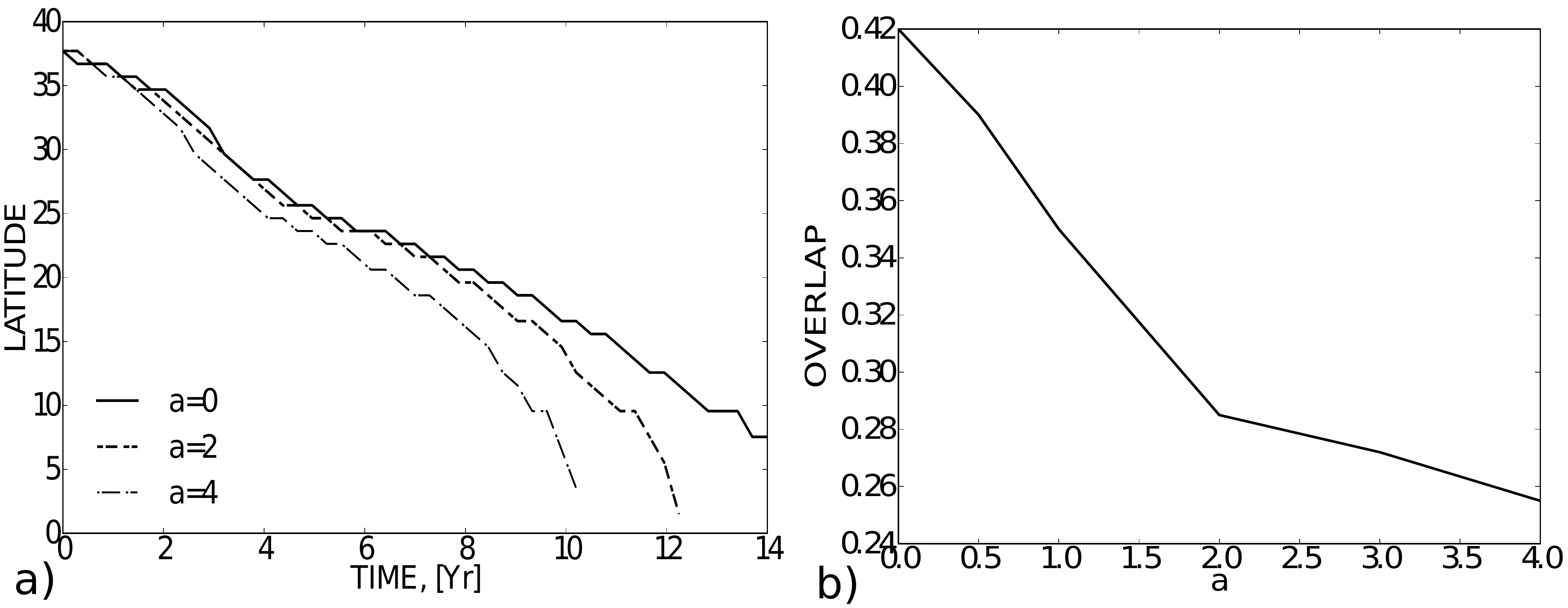} \caption{\label{fig:overlap}Panel a) The latitudinal location of maxima of
the toroidal magnetic field flux in the subsurface shear layer for
the model with $a=0,2,4$; b) the overlap time (relative to the cycle
length) between the subsequent cycles as a function of the anisotropy
parameter $a$. }
\end{figure}

Typical snapshots for the magnetic field and magnetic helicity distributions
in the solar convection zone are shown in Figure \ref{fig:prof}(b,c).
Similar to the benchmark models we see that the toroidal magnetic
field is concentrated to the boundaries of the convection zone. The
near surface magnetic field is weaker than the bottom field during
most of the dynamo cycle.

Figure \ref{fig:diag} shows the time -latitude diagram of the toroidal
magnetic field in the subsurface shear layer $r=0.92R_{\odot}$ and
for the radial magnetic field at the top of the domain $r=0.99R_{\odot}$.
Their behavior is similar to the solar cycles. We see that the cycle
period decreases from 14 Yrs to 10 Yrs when $a$ increases from $0$
to 4. For $a=4$ the model has the total effective magnetic diffusivity
greater than $10^{12}$ cm$^{2}$/s in a large part of the convection
zone. Still, the dynamo model reproduces correctly the solar cycle
period and the patterns of the toroidal and poloidal magnetic field
evolution.

Another interesting feature which is demonstrated by Figure \ref{fig:diag}
is that the overlap between the cycles decreases when we increase
the anisotropy parameter $a$. We investigated this feature in details
for the distributed dynamo model without meridional circulation. To
quantify the overlap between the subsequent cycles we examine the
latitudinal drift of the toroidal flux maxima in the subsurface shear
layer at $r=0.92R_{\odot}$(see Figure \ref{fig:overlap}a). We restrict
our consideration to the latitudinal range $0-40^{\circ}$, and compute
the relative overlap between the curves that belong to subsequent
cycles. Figure \ref{fig:overlap}b shows the results for the models
with $0\le a\le4$. The relative overlap time between the subsequent
cycles decreases from $0.42$(5 Yrs) to $0.25$(2.5 Yrs) when $a$
spans from $0$ to $4$. The speed of the dynamo wave latitudinal
migration remains almost the same for all $a$. This means that the
dynamo wave is not transformed from the running to steady type with
the increase of $a$.

\section{Discussion and conclusion}

In this paper we have examined influence of anisotropic turbulence
in the solar convection zone on magnetic diffusivity and properties
of mean-field dynamo models, including the simplified benchmark model
\citep{2008A&A...483..949J} and the distributed-dynamo model with
the subsurface shear layer and a detailed formulation of the mean-electromotive
force \citep{pk11,pip2013ApJ,PK13}. To characterized
the anisotropy, we, similar to \citet{2005A&A...431..345R}, use parameter
$a$ derived as a relative difference between horizontal and vertical
turbulent RMS velocities.

The strength of the anisotropy parameter depends on the theoretical
models of turbulent flows. In the simple case of the mixing length
theory, it is restricted to the range $a=0-4$ ($a=0$ corresponds
to isotropic flows). {The nonlocal stellar convection theory
\citep{chan06} suggests $a\sim3.5$.} Numerical simulations of the
solar convection showed evidence for a stronger anisotropy, which
may reach $a\sim10$ (e.g., \citealt{miesch08,2013arXiv1301.1330G}).
This anisotropy results from the global rotation effect on convective
motions, and forms the ``banana''-like giant convection cells in
the meridional direction. It is important to note that in the simulations
such anisotropic convection becomes visible even in the subsurface
shear layer below $0.95R_{\odot}$ where the analytical mixing-length
estimations of the mean-field diffusivity coefficients show a small
anisotropy. However, the current simulations still do not accurately
reproduce the convection zone dynamics. Therefore, we considered the
anisotropic parameter $a$ as a free parameter. The analytical calculations
presented in Appendix showed that the effect of turbulent mixing in
the horizontal direction is quenched by rotation while the effective
structure of the given anisotropic difusivity tensor remains the same
as in the case of slow rotation. 

{Study the benchmark models suggests that the dynamo threshold
parameters of the models change only a little with increase of the
anisotropy parameter $a$. This could be expected from analysis given
by \citet{k02}. The interesting new finding is that the dynamo threshold
can decrease with the increase of $a$. This is found for the models
with the odd number of circulation cell stacked along the radius.
For the case of the zero and double-cell circulation we find the opposite
behaviour. }

{In the benchmark models the equatorial drift of the toroidal
magnetic field depends on the amplitude and the type of meridional
circulation. The model B, which has no meridional circulation, does
not have equatorial branch of the toroidal magnetic field in the convection
zone. This is also due to the absence of the subsurface shear layer.
The detailed solar dynamo model, which was discussed in the subsection
2.3, has the equatorial migration of the toroidal magnetic field in
the upper part of the convection zone. This effect is induced by the
subsurface shear layer. The effect of the subsurface shear on the
large-scale dynamo was suggested earlier by \citet{bran05}. We note
the particular role of the boundary condition (Eq.\ref{eq:tor-vac})
for the subsurface shear to be feasible for the large-scale dynamo.
This condition results to penetration of the toroidal field to the
surface, which increase the efficiency of the subsurface shear layer
\citep{pk11apjl}. Thus, the model can produce the equatorial drift
of the toroidal field in the upper part of the convection zone (above
0.9$R_{\odot}$) for rather small equatorial turbulent pumping, which
operates in the bulk of the convection zone with amplitude less than
1 m/s (see, \citealp{pk11}). This is different to the model sugested
by \citet{2006AN....327..884K}, who employed, in addition to subsurface
shear, the strong turbulent pumping effects in the bulk of the convection
zone. The reader can look for the detailed discussion effect of subsurface
shear layer in above cited papers. }

{In the flux transport models C2 and C3 the equatorial branch
of the toroidal field is induced by the equatorial parts of the meridional
circulation cells. Thus, the model C2 has equatorial migration of
the toroidal magnetic field in the middle of the convection zone (see
more details in \citealt{PK13}). The model C3, with the triple-cell
circulation has in addition the equatorial migration near the bottom
of the convection zone. The physical interpretation of these types
of models suggests that poloidal field at the surface is produced
directly from the buoyant flux tubes that come from the bottom of
the convection zone. In this case the model with double-cell circulation
is not solar-like. The model with the triple-cell circulation has
a qualitatve agreement with observations, though the polar reversal
of the radial magnetic field, $\overline{B}_{r}$, occurs about 5
years ahead of the maximum of the toroidal magnetic field, $\overline{B}_{\phi}$,
in equatorial region. This model has the dynamo period as twice as
large compare to the solar cycle in the case $a=4$. For the case
$a=0$ model C3 has the larger dynamo period about 30 Yrs, though
the phase relation between activity of $\overline{B}_{r}$ and $\overline{B}_{\phi}$
is in a better agreement with observations.}

{We made addtional calculations for the distributed solar dynamo
models like that discussed in subsection 2.3 including the effect
of multi-cells circulation. For the case of the double-cell circulation
we reproduced our results from the previous paper (see, \citealp{PK13}).
We confirm the effect of the decreasing overlap between the subsequent
cycles with the increasing parameter of the anisotropy $a$. The case
of the triple-cell circulation is found to be similar to the model
C3. However in this case the correct phase relation} {between
activity of $\overline{B}_{r}$ and $\overline{B}_{\phi}$ holds only
for the upper part of the convection zone, where similar to the model
with double cell circulation we have the equatorial drift of the toroidal
magnetic field because of equatorward flow of the meridional circulation.
The results of that model are similar to results of the numerical
simulation reported by \citet{kap2012}. Note, that differential rotation
profile in their paper is different from the solar case.}

With the effects of anisotropy the total magnetic diffusivity reaches
values of about $10^{12}$cm$^{2}$/s in the layer $0.85-0.95R_{\odot}$,
which are consistent with estimates from the sunspot decay rate \citep{1993A&A...274..521M}
and the cross-helicity observations \citep{2011SoPh..269....3R}.
We found that the anisotropy affects the overlap time between the
subsequent cycles. When the anisotropy is larger, the overlap is smaller.
This effect is related to stronger concentration of the toroidal magnetic
field near the bottom boundary of the convection zone due to anisotropy,
and also as a consequence of a faster migration of the dynamo wave
in the subsurface shear layer. In general, our model is consistent
with the paradigm of solar dynamo operating in the bulk of the convection
zone and ``shaped'' in the subsurface roatational shear layer.

\section{Appendix A}

\setcounter{equation}{0}

\renewcommand{\theequation}{A\arabic{equation}}

To compute the electromotive force $\mathcal{E}$ for the anisotropic
MHD turbulence we write equations (\ref{induc}) and (\ref{fluc})
in the Fourier space:

\begin{eqnarray}
\left(\frac{\partial}{\partial t}+\eta z'^{2}\right)\hat{b}_{j} & \!=\! & iz'_{l}\int\left\{ \widehat{u}_{j}(\mathbf{z'-q)}B_{l}\left(\mathbf{q}\right)-\widehat{u}_{l}\left(\mathbf{z'-q)}B_{j}\left(\mathbf{q}\right)\right.\right\} \mathd\mathbf{q}+\widehat{\mathbf{\mathfrak{G}}}_{j}.\label{induc2}\\
\left(\frac{\partial}{\partial t}+\nu z^{2}\right)\hat{u}_{i} & \!=\! & \!\hat{f}_{i}+\hat{\mathfrak{F}}_{i}-2(\Omega\hat{z})(\hat{z}\times\hat{u})_{i}+\frac{i\pi_{if}}{\mu}z_{l}\int\{\widehat{b}_{l}(\mathbf{z-q)}B_{f}(\mathbf{q})+\widehat{b}_{f}(\mathbf{z-q)}B_{l}(q)\}\mathd\mathbf{q},\label{navie2}
\end{eqnarray}
where the turbulent pressure was excluded from (Eq.\ref{fluc}) by
convolution with tensor $\pi_{ij}(\mathbf{z)}=\delta_{ij}-\hat{z}_{i}\hat{z}_{j}$,
where $\delta_{ij}$ is the Kronecker symbol, and $\widehat{\mathbf{z}}$
is a unit wave vector. The equations for the second-order moments
that make contributions to the mean electromotive force can be found
directly from Eqs.(\ref{induc2}, \ref{navie2}). We consider the
high Reynolds number limit and discard the microscopic diffusion terms.
Using the $\tau$-approximation, in which the third order products
of the fluctuating fields are approximated by the corresponding relaxation
terms of the second-order contributions (see, \citealp{rad-kle-rog,kle-rog:04a,2005PhR...417....1B,pi08Gafd,garr2011}),
we arrive to

\begin{eqnarray}
\frac{\hat{\varkappa}_{ij}(\mathbf{z},\mathbf{z}')}{\tau^{*}\left(z\right)}+2(\Omega\hat{z})\varepsilon_{inm}\hat{z}_{n}\hat{\varkappa}_{mj}(\mathbf{z},\mathbf{z}') & = & iz'_{l}\int\left\{ \hat{v}_{ij}(\mathbf{z},\mathbf{z}'-\mathbf{q})\overline{B}_{l}(\mathbf{q})-\hat{v}_{il}(\mathbf{z},\mathbf{z}'-\mathbf{q})\overline{B}_{j}(\mathbf{q})\right\} \mathd\mathbf{q}\label{eq:kap}\\
 & + & \frac{i}{\mu}z_{l}\pi_{if}\int\hat{m}_{lj}(\mathbf{z}-\mathbf{q},\mathbf{z}')\overline{B}_{f}(\mathbf{q})\mathd\mathbf{q}\nonumber \\
 & + & \frac{i}{\mu}z_{l}\pi_{if}\int\hat{m}_{fj}(\mathbf{z}-\mathbf{q},\mathbf{z}')\overline{B}_{l}(\mathbf{q})\mathd\mathbf{q},\nonumber 
\end{eqnarray}

\begin{eqnarray}
\frac{\hat{v}_{ij}(\mathbf{z},\mathbf{z}')}{\tau^{*}\left(z\right)} & = & \frac{v_{ij}^{(0)}}{\tau^{*}\left(z\right)}-2(\Omega\hat{z})\varepsilon_{inm}\hat{z}_{n}\hat{v}_{mj}(\mathbf{z},\mathbf{z}')-2(\Omega\hat{z}')\varepsilon_{jnm}\widehat{z'}_{n}\hat{v}_{im}(\mathbf{z},\mathbf{z}')\label{eq:str}\\
 & + & \frac{i\pi_{if}}{\mu}z_{l}\int\{\hat{\varkappa}_{lj}(\mathbf{z}',\mathbf{z}-\mathbf{q})\overline{B}_{f}(\mathbf{q})+\hat{\varkappa}_{fj}(\mathbf{z}',\mathbf{z}-\mathbf{q})\overline{B}_{l}(q)\}\mathd\mathbf{q}\nonumber \\
 & + & \frac{i\pi_{jf}}{\mu}z'_{l}\int\{\hat{\varkappa}_{il}(\mathbf{z},\mathbf{z}'-\mathbf{q})\overline{B}_{f}(\mathbf{q})+\hat{\varkappa}_{if}(\mathbf{z},\mathbf{z}'-\mathbf{q})\overline{B}_{l}(q)]\mathd\mathbf{q},\nonumber \\
\frac{\hat{m}_{ij}(\mathbf{z},\mathbf{z}')}{\tau^{*}\left(z\right)} & = & iz'_{l}\int\{\hat{\varkappa}_{jl}(\mathbf{z}'-\mathbf{q},\mathbf{z})\left.\overline{B}_{l}(\mathbf{q})-\hat{\varkappa}_{li}(\mathbf{z}'-\mathbf{q},\mathbf{z})\overline{B}_{j}(\mathbf{q})\right\} \mathd\mathbf{q}\label{eq:max}\\
 & + & iz_{l}\int\{\hat{\varkappa}_{ij}(\mathbf{z}-\mathbf{q},\mathbf{z}')\left.\overline{B}_{l}(\mathbf{q})-\hat{\varkappa}_{lj}(\mathbf{z}-\mathbf{q},\mathbf{z}')\overline{B}_{i}(\mathbf{q})\right\} \mathd\mathbf{q}\nonumber \\
 & + & \frac{m_{ij}^{(0)}(\mathbf{z},\mathbf{z}')}{\tau^{*}\left(z\right)},\nonumber 
\end{eqnarray}
where we introduced the ensemble averages $\hat{v}_{ij}(\mathbf{z},\mathbf{z}')=\overline{\mathbf{u}_{i}\left(\mathbf{z}\right)\mathbf{u}_{j}\left(\mathbf{z}'\right)}$,
$\hat{\varkappa}_{ij}(\mathbf{z},\mathbf{z}')=\overline{\mathbf{u}_{i}\left(\mathbf{z}\right)\mathbf{b}_{j}\left(\mathbf{z}'\right)}$,
$\hat{m}_{ij}(\mathbf{z},\mathbf{z}')=\overline{\mathbf{b}_{i}\left(\mathbf{z}\right)\mathbf{b}_{j}\left(\mathbf{z}'\right)}$;
the superscript $^{(0)}$ stands for the background state (when the
mean-field is absent) of these correlations. The reader can find a
comprehensive discussion of the $\tau$ -approximation in the above
cited papers. Furthermore, the contributions of the mean magnetic
field in the equation for the turbulent stresses, $\hat{v}_{ij}$,
will be neglected because they give nonlinear terms in the cross-helicity
tensor, $\hat{\varkappa}_{ij}$.

Next, we solve Eqs(\ref{eq:kap},\ref{eq:str},\ref{eq:max}) in a
linear approximation for the mean field $\overline{\mathbf{B}}$,
also neglecting effects of background magnetic fluctuations in the
further analysis. Thus, we obtain

\begin{eqnarray}
\hat{\varkappa}_{ij}(\mathbf{z},\mathbf{z}') & = & i\tau^{*}\left(z\right)z'_{l}D_{ip}(\mathbf{z})\int\left\{ \hat{v}_{pj}(\mathbf{z},\mathbf{z}'-\mathbf{q})\overline{B}_{l}(\mathbf{q})-\hat{v}_{pl}(\mathbf{z},\mathbf{z}'-\mathbf{q})\overline{B}_{j}(\mathbf{q})\right\} \mathd\mathbf{q}\label{eq:kappa1}\\
 & + & D_{ip}(\mathbf{z})\frac{iz_{l}\tau^{*}\left(z\right)}{\mu}\pi_{pf}\int\hat{m}_{lj}(\mathbf{z}-\mathbf{q},\mathbf{z}')\overline{B}_{f}(\mathbf{q})\mathd\mathbf{q}\nonumber \\
 & + & D_{ip}(\mathbf{z})\frac{iz_{l}\tau^{*}\left(z\right)}{\mu}\pi_{pf}\int\hat{m}_{fj}(\mathbf{z}-\mathbf{q},\mathbf{z}')\overline{B}_{l}(\mathbf{q})\mathd\mathbf{q},\nonumber 
\end{eqnarray}

\begin{eqnarray}
\hat{v}_{ij} & = & \mathcal{D}_{injm}(\mathbf{z},\mathbf{z}')v_{nm}^{(0)}\label{eq:secm1}\\
 & + & \mathcal{D}_{injm}(\mathbf{z},\mathbf{z}')\frac{i\tau^{*}\left(z\right)\pi_{nf}}{\mu}z_{l}\int\{\hat{\varkappa}_{lm}(\mathbf{z}',\mathbf{z}-\mathbf{q})\overline{B}_{f}(\mathbf{q})+\hat{\varkappa}_{fm}(\mathbf{z}',\mathbf{z}-\mathbf{q})\overline{B}_{l}(q)\}\mathd\mathbf{q}\nonumber \\
 & + & \mathcal{D}_{injm}(\mathbf{z},\mathbf{z}')\frac{i\tau^{*}\left(z\right)\pi_{mf}}{\mu}z'_{l}\int\{\hat{\varkappa}_{nl}(\mathbf{z},\mathbf{z}'-\mathbf{q})\overline{B}_{f}(\mathbf{q})+\hat{\varkappa}_{if}(\mathbf{z},\mathbf{z}'-\mathbf{q})\overline{B}_{l}(q)]\mathd\mathbf{q},\nonumber 
\end{eqnarray}
where, 
\begin{eqnarray*}
D_{ip}(z) & = & \frac{\delta_{ip}+E_{ip}}{1+\psi_{\Omega}^{2}},\\
\mathcal{D}_{injm}(\mathbf{z},\mathbf{z}') & = & \left\{ \frac{\delta_{if}\hat{N}+2E_{if}}{\hat{N}^{2}+4\psi_{\Omega}^{2}}\right\} (\delta_{jm}(\delta_{fn}+E_{fn})-\delta_{fn}\tilde{E}{}_{jm})\\
E_{nk} & = & \frac{2(\Omega\cdot\mathbf{z})\tau}{\mathbf{z}^{2}}z_{p}\varepsilon_{nkp},\,\tilde{E}{}_{ml}=\frac{2(\Omega\cdot\mathbf{z}')\tau}{\mathbf{z}'^{2}}z'_{p}\varepsilon_{mlp},\\
\hat{N} & = & (1-\psi_{\Omega}^{2}+\tilde{\psi}{}_{\Omega}^{2}),\,\tilde{\psi}_{\Omega}=\frac{2(\Omega\cdot\mathbf{z}')\tau^{*}\left(z\right)}{|\mathbf{z}'|},\,\psi_{\Omega}=\frac{2(\Omega\cdot\mathbf{z})\tau^{*}\left(z\right)}{|\mathbf{z}|}
\end{eqnarray*}
To proceed further, we have to introduce some conventions and notations
that are widely used in the literature. The double Fourier transformation
of an ensemble average of two fluctuating quantities, say $f$ and
$g$, taken at equal times and at the different positions $\mathbf{x},\hspace{0.25em}\mathbf{x}'$,
is given by 
\begin{equation}
\left\langle f\left(\mathbf{x}\right)g\left(\mathbf{x}'\right)\right\rangle =\int\int\left\langle \hat{f}\left(\mathbf{z}\right)\hat{g}\left(\mathbf{z}'\right)\right\rangle e^{i\,\left(\mathbf{z}\cdot\mathbf{x}+\mathbf{z}'\cdot\mathbf{x}'\right)}\mathd^{3}\mathbf{z}\mathd^{3}\mathbf{z}'.\label{cor1}
\end{equation}
In the spirit of the general formalism of the two-scale approximation
{\citep{rob-saw}} we introduce ``fast'' and ``slow'' variables.
They are defined by the relative $\mathbf{r}=\mathbf{x}-\mathbf{x}'$
and the mean $\mathbf{R}=\frac{1}{2}\left(\mathbf{x}+\mathbf{x}'\right)$
coordinates, respectively. Then, eq. (\ref{cor1}) can be written
in the form 
\begin{eqnarray}
\left\langle f\left(\mathbf{x}\right)g\left(\mathbf{x}'\right)\right\rangle  & = & \int\int\left\langle \hat{f}\left(\mathbf{k}+\frac{1}{2}\mathbf{K}\right)\hat{g}\left(-\mathbf{k}+\frac{1}{2}\mathbf{K}\right)\right\rangle \mathe^{i\,\left(\mathbf{K}\cdot\mathbf{R}+\mathbf{k}\cdot\mathbf{r}\right)}\mathd^{3}\mathbf{K}\mathd^{3}\mathbf{k},\label{cor1b}
\end{eqnarray}
where we have introduced the wave vectors $\mathbf{k}=\frac{1}{2}\left(\mathbf{z}-\mathbf{z}'\right)$
and $\mathbf{K}=\mathbf{z}+\mathbf{z}'$. Then, following BS05, we
define the correlation function of $\widehat{\mathbf{f}}$ and $\widehat{\mathbf{g}}$
obtained from (\ref{cor1b}) by integration with respect to $\mathbf{K}$,
\begin{equation}
\Phi\left(\hat{f},\hat{g},\mathbf{k},\mathbf{R}\right)=\int\left\langle \hat{f}\left(\mathbf{k}+\frac{1}{2}\mathbf{K}\right)\hat{g}\left(-\mathbf{k}+\frac{1}{2}\mathbf{K}\right)\right\rangle \mathe^{i\,\left(\mathbf{K}\cdot\mathbf{R}\right)}\mathd^{3}\mathbf{K}.\label{eq:cor1a}
\end{equation}
For further convenience we define the second-order correlations of
velocity field and the cross-correlations of velocity and magnetic
fluctuations via 
\begin{eqnarray}
\hat{v}_{ij}\left(\mathbf{k},\mathbf{R}\right) & = & \Phi(\hat{u}_{i},\hat{u}_{j},\mathbf{k},\mathbf{R}),\left\langle u^{2}\right\rangle \left(\mathbf{R}\right)=\int\hat{v}_{ii}\left(\mathbf{k},\mathbf{R}\right)\mathd^{3}\mathbf{k},\\
\hat{\varkappa}_{ij}\left(\mathbf{k},\mathbf{R}\right) & = & \Phi(\hat{u}_{i},\hat{b}_{j},\mathbf{k},\mathbf{R}),\mathcal{E}_{i}\left(\mathbf{R}\right)=\varepsilon_{ijk}\int\hat{\varkappa}_{jk}\left(\mathbf{k},\mathbf{R}\right)\mathd^{3}\mathbf{k}.
\end{eqnarray}
We now return to equations (\ref{eq:kappa1}) and (\ref{eq:secm1}).
As the first step, we perform the Taylor expansion with respect to
the ``slow'' variables, and take the Fourier transform, (\ref{eq:cor1a}).
The details of this procedure can be found in \citep{2005PhR...417....1B}.
In result we get the following equations for the second moments

\begin{eqnarray}
\hat{\varkappa}_{ij} & = & -i\tau^{*}D_{if}\left(\mathbf{B}\cdot\mathbf{k}\right)\hat{v}_{fj}-\tau^{*}D_{if}\hat{v}_{fl}\overline{B}_{j,l}\label{eq:solF}\\
\hat{v}_{ij} & = & T_{ijnm}^{(0)}\hat{v}_{nm}^{(0)},\nonumber 
\end{eqnarray}
where

\begin{eqnarray*}
T_{ijnm}^{\left(0\right)} & = & \delta_{in}\delta_{jm}+\frac{\psi_{\Omega}\hat{k}_{p}}{M}\left(\varepsilon_{inp}\delta_{jm}+\varepsilon_{jmp}\delta_{in}\right)\\
 & - & \frac{\psi_{\Omega}^{2}}{M}\left(\delta_{ij}\pi_{nm}-\delta_{nm}\hat{k}_{i}\hat{k}_{j}+\delta_{im}\hat{k}_{n}\hat{k}_{j}+\delta_{nj}\hat{k}_{i}\hat{k}_{m}-2\delta_{n\left[i\right.}\delta_{\left.j\right]m}\right),\\
M & = & 1+4\psi_{\Omega}^{2}
\end{eqnarray*}
The statistical properties of the background fluctuations are described
by the spectral tensor (see, e.g. \citealt{2005A&A...431..345R}):
\begin{eqnarray}
\hat{v}_{ij}^{(0)} & = & \left\{ \pi_{ij}\left(\mathbf{k}\right)\frac{E\left(k,\mathbf{R}\right)}{8\pi k^{2}}+\frac{E_{1}\left(k,\mathbf{R}\right)}{8\pi k^{2}}\left(\pi_{ij}\left(\mathbf{g}\right)\pi_{nm}\left(\mathbf{g}\right)-\pi_{in}\left(\mathbf{g}\right)\pi_{jm}\left(\mathbf{g}\right)\right)\hat{k}_{n}\hat{k}_{m}\right\} ,\label{eq:spectr1-1}
\end{eqnarray}
where, the spectral functions $E(k,\mathbf{R})$ and $E_{1}(k,\mathbf{R})$
define the intensity of the velocity fluctuations in the radial and
horizonthal directions, and $\mathbf{g}$ is a unit vector in the
direction of anisotropy. Following the conventions given by \citet{2005A&A...431..345R}
we write 
\begin{eqnarray}
\left\langle u_{r}^{(0)2}\right\rangle  & = & \frac{1}{3}\int\frac{E\left(k,\mathbf{R}\right)}{4\pi k^{2}}\mathd^{3}\mathbf{k}\label{eq:spectr3-1}\\
\left\langle u_{h}^{(0)2}\right\rangle -2\left\langle u_{r}^{(0)2}\right\rangle  & = & \frac{1}{3}\int\frac{E_{1}\left(k,\mathbf{R}\right)}{4\pi k^{2}}\mathd^{3}\mathbf{k}
\end{eqnarray}
and introduce a non-dimensional anisotropy parameter $a$: 
\begin{equation}
a=\frac{\left(\left\langle u_{h}^{(0)2}\right\rangle -2\left\langle u_{r}^{(0)2}\right\rangle \right)}{\left\langle u_{r}^{(0)2}\right\rangle }.\label{eq:anis}
\end{equation}
Note, that $a\ge-1$ because of the Bochner's theorem (see, e.g.,
\citealt{1975mit..bookR....M}). To integrate Eqs(\ref{eq:solF})
in the k-space we apply the Kolomogorov spectra for $E\left(k,\mathbf{R}\right)=-\left\langle u^{(0)2}\right\rangle d\overline{\tau}\left(k\right)/dk$,
$\overline{\tau}\left(k\right)={\displaystyle \left(\frac{k}{k_{0}}\right)^{1-q}}$and
$\tau^{*}=2\tau_{c}\overline{\tau}\left(k\right)$ with $q=5/3$,
$k_{0}=1/\ell_{0}$, $\tau_{c}=\ell_{0}/\sqrt{\left\langle u^{(0)2}\right\rangle }$
(see, e.g., \citealp{rad-kle-rog,kle-rog:04b}). We assume that $a$
is a constant over $k$. This approximation can be refined in further
applications.

The isotropic part of magnetic diffusivity in rotating turbulent media
was derived by \citet{pi08Gafd}, and is not reproduced here. The
effect of the anisotropic mixing and the Coriolis force on magnetic
diffusivity is given by 
\begin{eqnarray}
\mathcal{E}_{i}^{(A)} & = & a\left\langle u^{(0)2}\right\rangle \tau_{c}\left\{ \varepsilon_{\mathrm{ijm}}(f_{5}(\mathbf{e}\cdot\mathbf{g})^{2}+f_{1})B_{\mathrm{j,m}}+\varepsilon_{\mathrm{ijm}}B_{\mathrm{j,f}}\left((f_{5}+f_{4}(\mathbf{e}\cdot\mathbf{g})^{2})e_{f}e_{m}-f_{9}g_{f}e_{m}\mathrm{(\mathbf{e}\cdot\mathbf{g})}\right)\right.\label{eq:result}\\
 & + & \varepsilon_{\mathrm{ijm}}\left(f_{4}e_{f}e_{l}\mathrm{\mathrm{(\mathbf{e}\cdot\mathbf{g})}}+f_{6}e_{f}g_{l}-f_{3}g_{f}e_{l}\right)e_{j}g_{m}B_{\mathrm{f,l}}\nonumber \\
 & + & e_{j}g_{m}\varepsilon_{\mathrm{jmf}}(f_{3}g_{i}e_{l}-f_{4}e_{i}e_{l}(\mathbf{e}\cdot\mathbf{g})-f_{6}e_{i}g_{l})B_{\mathrm{f,l}}\nonumber \\
 & + & g_{j}\varepsilon_{\mathrm{jmf}}(f_{3}e_{i}(\mathbf{e}\cdot\mathbf{g})+f_{7}g_{i})B_{\mathrm{m,f}}+g_{j}\varepsilon_{\mathrm{ijm}}(f_{8}e_{f}(\mathbf{e}\cdot\mathbf{g})+f_{10}g_{f})B_{\mathrm{m,f}}\nonumber \\
 & + & g_{j}\varepsilon_{\mathrm{ijm}}\left(f_{3}e_{f}\mathrm{(\mathbf{e}\cdot\mathbf{g})}+f_{7}g_{f}\right)B_{\mathrm{f,m}}+f_{3}g_{f}e_{m}\varepsilon_{\mathrm{mfl}}\mathrm{(\mathbf{e}\cdot\mathbf{g})}B_{\mathrm{l,i}}+\nonumber \\
 & + & \left.f_{2}(g_{i}e_{l}-e_{i}g_{l})e_{j}g_{m}\varepsilon_{\mathrm{jmf}}B_{\mathrm{l,f}}\right\} +\dots\delta-effect,\nonumber 
\end{eqnarray}
where ${\displaystyle \mathbf{e}=\frac{\boldsymbol{\Omega}}{\Omega}}$
is the unit vector of angular velocity; and $f_{1-10}$ are functions
of the Coriolis number, $\Omega^{\star}=2\Omega_{0}\tau_{c}$. The
last term in Eq(\ref{eq:result}) means that there are addtional magnetic
field generation effects induced by the large-scale current and global
rotation, so-called $\delta$-effect, \citep{rad69}. For the background
hydrodynamic fluctuation we found no $\delta$-effect in the direction
of large-scale magnetic field. The other terms of the $\delta$-effect
can be less important for the solar type dynamo, and we skip them
from our consideration.

For the case of the slow rotation, taking the Taylor expansions of
$f_{1-10}$ about small $\Omega^{\star}$ we find that $f_{1,10}=\frac{1}{6}$,
and the others functions are order of $O\left(\Omega^{\star2}\right)$.
This reduces the Eq.(\ref{eq:result}) to 
\begin{equation}
\boldsymbol{\mathcal{E}}=-\frac{a}{2}\eta_{T}\left(\boldsymbol{\nabla}-\mathbf{g}\left(\mathbf{g}\cdot\nabla\right)\right)\times\overline{\mathbf{B}},\label{eq:simplE}
\end{equation}
where ${\displaystyle \eta_{T}=\frac{\tau_{c}}{3}\left\langle u^{(0)2}\right\rangle }$.

Despite a complicated form of Eq(\ref{eq:result}) only certain combinations
of $f_{1-10}$ are important in applications. For example, in the
case of the spherical geometry and the axisymmetric magnetic field
$\bar{\mathbf{B}}=\mathbf{e}_{\phi}B+\nabla\times\frac{A\mathbf{e}_{\phi}}{r\sin\theta}$,
where $B(r,\theta,t)$ is the azimuthal component, $A(r,\theta,t)$
is proportional to the azimuthal component of the vector potential,
we find 
\begin{eqnarray}
\mathcal{E}_{r} & = & \eta_{T}a\left\{ \frac{\phi_{1}}{r}\frac{\partial\sin\theta B}{\partial\mu}+\frac{\phi_{2}}{r}\mu\sin\theta B\right\} ,\nonumber \\
r\mathcal{E}_{\theta} & = & \eta_{T}a\left(\phi_{3}+\phi_{2}\mu^{2}\right)B,\label{eqs:emfA}\\
r\sin\theta\mathcal{E}_{\phi} & = & \eta_{T}a\left\{ \frac{\phi_{1}\sin^{2}\theta}{r^{2}}\frac{\partial^{2}A}{\partial\mu^{2}}+\frac{\phi_{1}}{r}\frac{\partial A}{\partial r}\right\} ,\nonumber 
\end{eqnarray}
where, $\mu=\cos\theta$ , $\phi_{1}=f_{1}+f_{3}+f_{5}+f_{7}$, $\phi_{2}=f_{9}+f_{6}-f_{3}+f_{2}$
, $\phi_{3}=f_{10}+f_{7}-f_{2}-f_{6}$ and 
\begin{eqnarray}
\phi_{1} & = & -\frac{1}{24\Omega^{\star2}}\left(2\log\left(1+4\Omega^{\star2}\right)+4\log\left(1+\Omega^{\star2}\right)+\right.\label{eq:phi1}\\
 &  & +\left.\left(1-4\Omega^{\star2}\right)\frac{\arctan\left(2\Omega^{\star}\right)}{\Omega^{\star}}+4\left(1-\Omega^{\star2}\right)\frac{\arctan\left(\Omega^{\star}\right)}{\Omega^{\star}}-6\right),\nonumber \\
\phi_{2} & = & -\frac{1}{24\Omega^{\star2}}\left(4\log\left(1+4\Omega^{\star2}\right)+8\log\left(1+\Omega^{\star2}\right)+\right.\label{eq:phi2}\\
 & + & \left.\left(3-4\Omega^{\star2}\right)\frac{\arctan\left(2\Omega^{\star}\right)}{\Omega^{\star}}+4\left(3-\Omega^{\star2}\right)\frac{\arctan\left(\Omega^{\star}\right)}{\Omega^{\star}}-18\right),\nonumber \\
\phi_{3} & = & \frac{1}{12\Omega^{\star2}}\left(\log\left(1+4\Omega^{\star2}\right)+2\log\left(1+\Omega^{\star2}\right)+\frac{\arctan\left(2\Omega^{\star}\right)}{\Omega^{\star}}+4\frac{\arctan\left(\Omega^{\star}\right)}{\Omega^{\star}}-6\right).\label{eq:phi3}
\end{eqnarray}
Note, that in the case of the non-small Coriolis number the only significant
effect is expressed with the terms with factor $\phi_{2}$. These
terms appears in the spherical geometry due to the anisotropy. The
functions $\phi_{1-3}$ describe a quenching of the given anisotropy
effect because, $\phi_{1,2}\sim\Omega^{\star-1}$ and $\phi_{3}\sim\log\left(\Omega^{\star}\right)/\Omega^{\star2}$
for $\Omega^{\star}\gg1$.


\end{document}